\documentclass[prl,amssymb,twocolumn,superscriptaddress]{revtex4}

\usepackage{graphicx}

\begin{document}

\title{Frequency control of photonic crystal membrane resonators by mono-layer deposition}

\author{Stefan Strauf} \email[corresponding author: ]{strauf@physics.ucsb.edu}
\affiliation{Physics Department, University of California Santa Barbara, CA 93106, Santa Barbara, USA}

\author{Matthew T. Rakher}
\affiliation{Physics Department, University of California Santa Barbara, CA 93106, Santa Barbara, USA}

\author{Itai Carmeli}
\affiliation{Physics Department, University of California Santa Barbara, CA 93106, Santa Barbara, USA}

\author{Kevin Hennessy}
\affiliation{ECE Department, University of California Santa Barbara, CA 93106, Santa Barbara, USA}

\author{Cedrik Meier}
\affiliation{ECE Department, University of California Santa Barbara, CA 93106, Santa Barbara, USA}

\author{Antonio Badolato}
\affiliation{ECE Department, University of California Santa Barbara, CA 93106, Santa Barbara, USA}

\author{Michiel J.A. DeDood}
\affiliation{Physics Department, University of California Santa Barbara, CA 93106, Santa Barbara, USA}

\author{Pierre M. Petroff}
\affiliation{ECE Department, University of California Santa Barbara, CA 93106, Santa Barbara, USA}
\affiliation{Materials Department, University of California Santa Barbara, CA 93106, Santa Barbara, USA}

\author{Evelyn L. Hu}
\affiliation{ECE Department, University of California Santa Barbara, CA 93106, Santa Barbara, USA}
\affiliation{Materials Department, University of California Santa Barbara, CA 93106, Santa Barbara, USA}

\author{Beth Gwynn}
\affiliation{Physics Department, University of California Santa Barbara, CA 93106, Santa Barbara, USA}

\author{Dirk Bouwmeester}
\affiliation{Physics Department, University of California Santa Barbara, CA 93106, Santa Barbara, USA}

\begin{abstract}
We study the response of GaAs photonic crystal membrane resonators to thin film deposition. Slow spectral shifts of the
cavity mode of several nanometers are observed at low temperatures, caused by cryo-gettering of background molecules.
Heating the membrane resets the drift and shielding will prevent drift altogether. In order to explore the drift as a
tool to detect surface layers, or to intentionally shift the cavity resonance frequency, we studied the effect of
self-assembled monolayers of polypeptide molecules attached to the membranes. The 2 nm thick monolayers lead to a
discrete step in the resonance frequency and partially passivate the surface.
\end{abstract}

\maketitle

Photonic crystal membrane microcavities (PCM) are promising candidates for applications ranging from quantum and
classical communication \cite{Bouwmeester:Book}, to microlasers \cite{Painter:99, Park:Science04} and sensing devices
\cite{Loncar:APL03, Chow:OL04}. Due to their ultra small mode volumes \cite{Vahala:Nature2003} and large surface to
volume ratio, the PCM resonant frequency is highly sensitive to its environment. While this sensitivity may be
exploited for novel sensing applications, it complicates solid-state cavity quantum electrodynamic (QED) experiments
that depend on a precise resonance condition between a cavity mode and an embedded single quantum dot (QD)
\cite{Michler:Science00, Vuckovic:APL03, Reithmeier:Nature04, Yoshie:Nature04}, single atom \cite{Vuckovic:PRE02} or
single impurity \cite{Strauf:PRL02}. This paper describes a slow red-shift of the PCM mode emission frequency that can
occur at low operation temperatures. We ascribe this shift to molecular condensation on the PCM surface. We further
describe methods used to fully curtail the drift, and in addition, we report on the first demonstration of a controlled
red-shift of the PCM mode by the adsorption of a self-assembled monolayer (SAM) of polypeptide molecules.\\
\indent We are particularly interested in PCM devices operated at low temperatures in such a way that embedded QDs
display a discrete energy spectrum. As a model system a square-lattice PCM geometry with one missing air hole (S1) has
been chosen, which is known to confine the fundamental mode in the proximity of the air-semiconductor interface
\cite{Kevin:APL05}. A single layer of self-assembled InAs QDs was embedded in the 180 nm thick GaAs membrane and emits
around 950 nm \cite{Garcia:APL98} under non-resonant laser excitation at 780 nm. Figure 1a shows a spectrum of the
fundamental cavity mode taken at pump powers of 15 $\mu$W, which has been recorded with a micro photoluminescence
(micro-PL) setup \cite{Brian:PRL05}.
\begin{figure}[tb!]
\includegraphics[width=72mm]{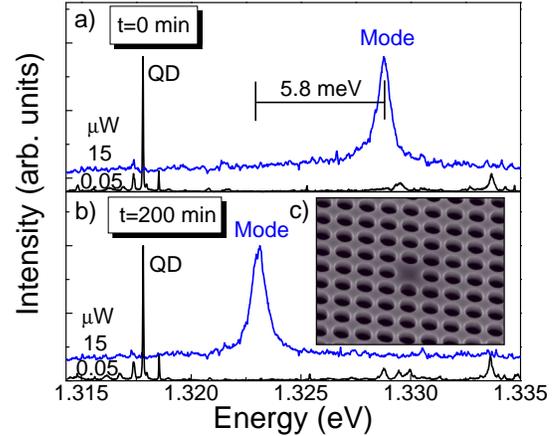}
\caption{Low temperature (4.2~K) micro-PL spectra of a PCM shortly after cool-down (a) and 200~min later (b). High pump
power spectra ($15~\mu$W, blue) show the cavity mode and low pump power spectra (0.05~$\mu$W, black) show a single QD
emitting at 1.318~eV. c) Scanning electron micrograph of a PCM with a lattice constant $a$ of 290 nm and hole radius
$r$ of 110 nm around the defect region.}
\end{figure}
At these excitation conditions the cavity mode is clearly visible at 1.3293 eV with a quality factor (Q-factor) of
1900. The 2~$\mu$m diameter laser excitation spot has to be positioned with an accuracy of $\pm 0.5~\mu$m with respect
to the cavity defect region (Fig. 1c), demonstrating the strongly localized character of the mode. Individual QD
transitions are visible under low pump power excitation of 50 nW, which have been identified by their pronounced
antibunching signature (not shown). Another set of two spectra was taken 200 min later as shown in Fig.~1b. While the
single QD emission energy at 1.31780 $\pm~2\cdot10^{-5}$~eV did not change in the entire observation period, the cavity
mode has now red-shifted by 5.8 meV (4 nm) and has a slightly lower Q-factor of 1700. The energetically stable QD
emission indicates that no temperature drift or strain-induced modification of the
electronic states \cite{Nakaoka:APL04} occurred over time.\\
\indent The cavity mode energy shift as a function of time is shown in Fig. 2a.
\begin{figure}[tb!]
\includegraphics[width=72mm]{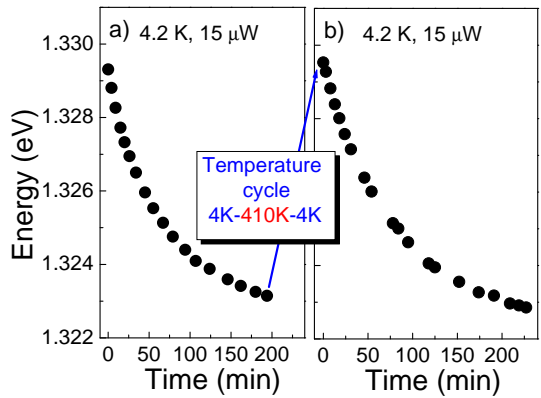}
\caption{(a) Drift of the cavity mode energy with time. Cycling the temperature from 4~K to 410~K and back to 4~K
resets the mode back to 1.3295~eV and the drift starts again as shown in Fig.2(b). Data are taken at 4.2~K and $15~\mu$W pump power.}
\end{figure}
The observed red-shift slows down and saturates after a few hours. Measurements on different samples show that the
drift is mostly independent of the actual $r/a$ ratio of the S1 cavities. It is known that chemical wet etching of the
PCM structures in HF and selective removal of a self-formed native oxide will result in a systematic blue shift of the
cavity mode \cite{Kevin:APL05}. Therefore, we believe that the measured red shift can be ascribed to material being
added (adsorbed) onto the same surface through cryo-gettering \cite{Oxford}. In confirmation of this hypothesis, we
have found that the red-shift of the cavity mode can be fully recovered by cycling the sample temperature from 4~K to
410~K and back
to 4~K (Fig. 2b), demonstrating that the thin film can be fully removed.\\
\indent  A more detailed study reveals that the temperature dependence of the mode shift is varying as shown in Fig. 3a.
\begin{figure}[tb!]
\includegraphics[width=72mm]{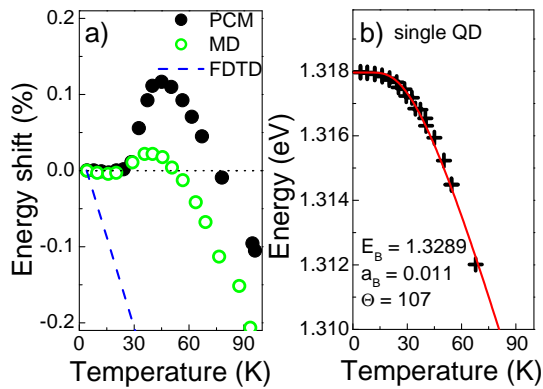}
\caption{(a) Temperature dependence of the mode emission energy for a PCM (solid dots) and a MD (open dots). Dashed
line: FDTD calculation of the mode shift with temperature for a PCM. (b) Temperature dependence of the single QD
transition energy (crosses). The red line is a fit according to the Bose-Einstein model. Mode energies are normalized
to the $T=4$~K values.}
\end{figure}
All temperature data have been recorded 5 hours after initial cool down to 4K in the regime where the mode drift over
time is saturated. With increasing temperature the mode energy is nearly constant between 4-30 K, blue shifts by 1 nm
between 30-50 K and is followed by a red shift of ~2 nm up to 100K (solid dots). In contrast, the single QD emission
energy (Fig. 3b, crosses) follows the expected temperature dependence of the semiconductor bandgap (Fig. 3b, red line),
according to the Bose-Einstein model \cite{Grilli:PRB1992} with parameters as given in the inset. This demonstrates
that the actual temperature readout of the Si-thermometer inside the cryostat is close to the real sample temperature
within the PCM defect region, where the single QD is located. For temperatures above 50 K the slope of the mode energy
follows the expected linear red-shift due to the temperature dependence of the effective refractive index $n(T)$, as
has been calculated by finite-difference time-domain simulations (dashed line) assuming that $n(T)$ of the GaAs
membrane changes with temperature according to \cite{nvonT}. Thus, the observed anomalous blue-shift at temperatures
around 40 K must be caused by an additional
effect and is attributed to a partial desorption and/or reconfiguration of the deposited film.\\
\indent By way of comparison, we studied the behavior of the whispering-gallery mode (WGM) of microdisk (MD) structures
with 5 $\mu$m diameter, which have been defined by optical lithography and transferred into the GaAs by a two-step wet
etch process based on HBr \cite{Michler:APL00}. These devices show a largely reduced blue-shift with increasing
temperature (Fig. 3a open dots) as well as a largely reduced frequency shift over time (Fig. 4, open green dots)
compared to PCM reference devices (Fig.4, blue dots). The WGMs of the MD structures have some evanescent coupling in
both lateral and vertical direction, but the PCM mode penetrates much further into the air-hole region \cite{Kevin:APL05}
and is therefore more susceptible to the environment.\\
\indent On the one hand the pronounced sensitivity of the PCM mode to the actual environmental conditions is promising
for chemical sensing applications. On the other hand this sensitivity might significantly complicate the analysis of
cavity-QED experiments utilizing temperature tuning in order to establish a resonance condition between the cavity mode
and the emission energy of an embedded single quantum dot \cite{Yoshie:Nature04}. In order to fully stop the mode
energy drift over time the PCM devices have been capped with a thin glass slide \cite{Film}. As a result, the energy
drift is now completely absent (Fig. 4 black dots).
\begin{figure}[tb!]
\includegraphics[width=80mm]{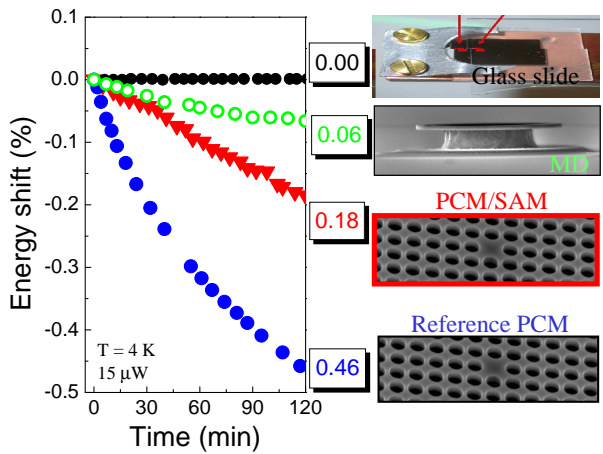}
\caption{Comparison of cavity mode red-shift over time for a reference PCM (blue dots), a SAM-coated PCM (red
triangles), a microdisk (open green dots) and a PCM capped with a thin glass slide (black dots). Mode energies are
normalized to the $t=0$~min values. Data are taken at 4.2~K and $15~\mu$W pump power.}
\end{figure}
While this approach is satisfactory for cavity-QED experiments at low temperatures, it would be highly desirable
to directly manipulate the PCM surface, allowing for selective sensing of chemical species and to interface with
functional molecules and/or colloidal QDs. To this end we linked polypeptide SAMs to the GaAs surface of the PCM. The
polypeptide molecule is composed of eight alternating (Ala-Aib) sequences with a Glutamic acid (Glu) attached to the
C-terminal of the peptide and is stable in $\alpha$-helix form \cite{Miura:Langmuir98}. The peptide chemically binds to
the GaAs surface by a carboxyl group located in the Glu amino acid. It was adsorbed to the GaAs surface by first
removing the surface oxide in dilute HF, and then immediately immersing it in 1 millimolar solution of the molecule
in absolute ethanol.\\
\indent The thickness of the monolayer was measured by ellipsometry to be $2.2 \pm 0.2$ nm, in agreement with the
calculated length of the polypeptide of 2.6 nm. This indicates that the molecules have formed a monolayer with the long
molecule axis almost perpendicular to the surface. Samples with an attached SAM and initially capped with a glass slide
show a red-shifted PCM mode by about 3-5 nm compared to reference samples without SAMs. This indicates that the SAM is
indeed attached to the GaAs surface and highlights the pronounced sensitivity of the S1 cavity mode's ability to sense
layers only 2 nm thick with a frequency response about 6-10 times larger than the full width at half maximum of the
cavity mode (0.5 nm). This sensitivity can be further increased by use of S1 cavities with
demonstrated Q-factors as high as 10000 \cite{Kevin:APL05}.\\
\indent Finally, we removed the glass slide from the SAM covered sample and found a largely reduced magnitude of the
cavity mode red-shift over time by up to a factor of three (Fig.4, red triangles) compared to untreated PCM reference
devices (Fig.4, blue dots). This demonstrates a partial surface passivation once a SAM is attached. It is furthermore
expected that the use of molecules with longer chains and thus larger average film thickness would give rise to a
further reduction of the mode shift over time.\\
\indent In summary, our experiments show how one may control the environment of a PCM to obtain either stable, stepped
or continuous tuning operation, each of which will be of interest in a variety of nanophotonic applications. We
demonstrated a new method to attach self-assembled monolayers to GaAs photonic crystal membrane cavities opening novel
possibilities for biofunctionalized photonic devices.

This research has been supported through NSF NIRT Grant no. 0304678 and DARPA Grant No. MDA 972-01-0027. One of the
authors, S. Strauf acknowledges support from the Max-Kade Foundation.


\begin{thebibliography}{30}
\bibitem{Oxford}
A similar drift of the PCM mode energy has been observed in four different optical cryostates (Oxford), although the
vacuum was better than $1 \times 10^{-6}$ mTorr.
\bibitem{Film}
The barely visible film gettered on the glass slide dissolves completely in acetone but not in deionized water,
indicating that the material is hydrocarbon related.
\end{thebibliography}
\end{document}